# An Overview of Planar Flow Casting of Thin Metallic Glasses and its Relation to Slot Coating of Liquid Films

Eric A. Theisen[1], Steven J. Weinstein[2]

1 - Metglas Inc., 440 Allied Dr., Conway, SC 29526, USA, Corresponding author: eric.theisen@metglas.com

2 - Department of Chemical Engineering, Rochester Institute of Technology, Rochester, NY, 14623, USA

**ABSTRACT**

Planar flow casting (PFC) is a method that can be used to make thin, long, and wide metallic alloy foils by extruding molten liquid through a thin and wide nozzle and immediately quenching on a moving roller. The quenching rates are high enough that amorphous metallic glasses may be formed which have many desirable properties for a wide variety of applications. This paper reviews how PFC processes were developed, examines the typical operability range of PFC, and reviews the defects that commonly form. The geometrical similarities between PFC and slot coating process are apparent, and this paper highlights differences between the operability ranges of both processes.

## 1 - Introduction

Amorphous alloys, often called metallic glasses, were initially discovered in the 1960's, when droplets of molten metal impinged on a quench substrate that yielded high cooling rates on the order of $10^5$ to $10^7$ C/s [1]. These were the first examples of metallic materials that had meta-stable morphologies where the kinetics of crystallization was avoided. Amorphous alloys are often referred to as thin metallic glasses (TMGs), as rapid cooling requires the solidifying material to be thin in the direction of heat extraction. Today, these TMGs are used in many applications that include high efficiency transformers and other energy efficient electrical conversion devices that reduce power generation around the globe [2].

The first amorphous metallic alloys were Gold-Silicon followed by other more industrial relevant compositions such as Iron-Phosphorus-Carbon [3] and others [4]. The high cooling rates resulted in high solidification velocities that avoided the morphological instability of constitutional undercooling that would typically result in dendritic structures. Generally, the composition of these alloys was close to a eutectic point in the phase diagram where it was easier to access a large undercooling. A general formula for amorphous alloys was found to be $M_{70-90} Y_{10-30} Z_{0-15}$ in atomic percentages where M is a transition metal (such as Iron, Cobalt, Nickel, etc.), Y is a non-metallic glass former (such as Boron, Carbon, Phosphorous, etc.), and Z is another metalloid (such as Beryllium, Aluminum, Silicon). To date, very few amorphous materials have been produced with deviations from this formula.

A variety of novel properties are associated with amorphous metals compared with their crystalline counterparts. The mechanical hardness is three to four times higher than conventional metals making them good candidates for new applications requiring surface wear resistance. The electrical conductivity is half of the value of a similar crystalline material leading to applications in resistance heating. The absence of grain boundaries in the amorphous material make metallic glasses ideal to examine mechanical toughness and fracture properties [5]. Amorphous alloys also exhibit superior corrosion resistance due to their compositional homogeneity [6].

One of the initial applications for TMGs was in high efficiency distribution transformer cores, first utilized in the US power grid in 1982 [7]. Transformer cores require wound laminations of magnetic materials, and the soft magnetic properties of amorphous foils (wide and long thin sheets) are ideal for this application. The lack of crystalline structures greatly reduces the hysteresis losses in the flux reversal process. Additionally, the thickness of amorphous foil is on the order of 25 microns due to the cooling rate requirements. This is 10-20 times thinner than a conventional rolled Silicon-Steel material, and the TMGs reduce the occurrence of eddy current losses.

Glass forming ability is enhanced when alloy compositions lie near deep eutectics in the phase diagram where elements such as Boron, Carbon, and Phosphorous, serve as melting point depressants. These additives lower the melting point below that of conventional metals and enable TMGs to be employed in metal joining applications [8]. Here the TMG is a filler material that is re-melted to bond the base materials. In most cases the amorphous foils are highly ductile and can be cut or stamped into preformed shapes making them very convenient for brazing. Such foils are favorable compared with more conventional metal joining materials in which powders imbedded in an organic binder are used in brazing. These organic binders are burned off in the brazing process and can cause porosity of the final joint. In this application the TMG foil is not a final product, but is used in an intermediate step—after brazing, the foil does transform to a crystalline state.

As new applications such as these have emerged, large scale production of foil TMGs has ensued. Commercial requirements necessitate that TMGs have uniform properties, such as thickness, width, composition, and of course structure. The planar flow casting (PFC) process was developed to meet these stringent requirements, and its history and attributes are the subject of this paper, whose organization is as follows. Section 2 discusses how the commercial PFC process was developed starting from initial benchtop trials. Section 3 provides typical PFC process conditions and highlights operational features that are common to most PFC machines. The uniformity of PFC foils and common defects are discussed in Section 4. Section 5 then further examines the physics of the PFC process and contrasts it with single-layer slot coating processes that have similar geometrical features.

**2 – Thin Metallic Glass Processing and the Development of Planar Flow Casting**

The initial droplet splat experiments cited in Section 1 were able to produce small flakes and shards of amorphous materials that were appropriate for laboratory scale studies. In these experiments, the droplets were impacted against a stationary quenching target, often a copper block (Figure 1a). The practical applications of metallic glasses motivated the development of continuous processes to produce such materials at larger scale. Early processes studied included casting onto a rotating substrate, a twin-roll co-rotating system and a belt style machine [9]. Thermal deformations of the twin roll or belt style method became problematic and the simplicity of the wheel method proved to be more readily adopted. The first commercially-produced amorphous foils were created using the method of free jet casting, in which a cylindrical jet of molten metal impinged on a moving substrate (Figure 1b). In free jet casting, the width of the foil is limited to only a few mm due to the stability of the molten metal puddle formed between the jet and the rotating wheel. Within its operability window, this method produced samples limited only by the supply of the jet and the ability to maintain a cool substrate. The early commercial applications took the narrow foils and wove them into larger sheets, and were used mainly used in magnetic shield applications.

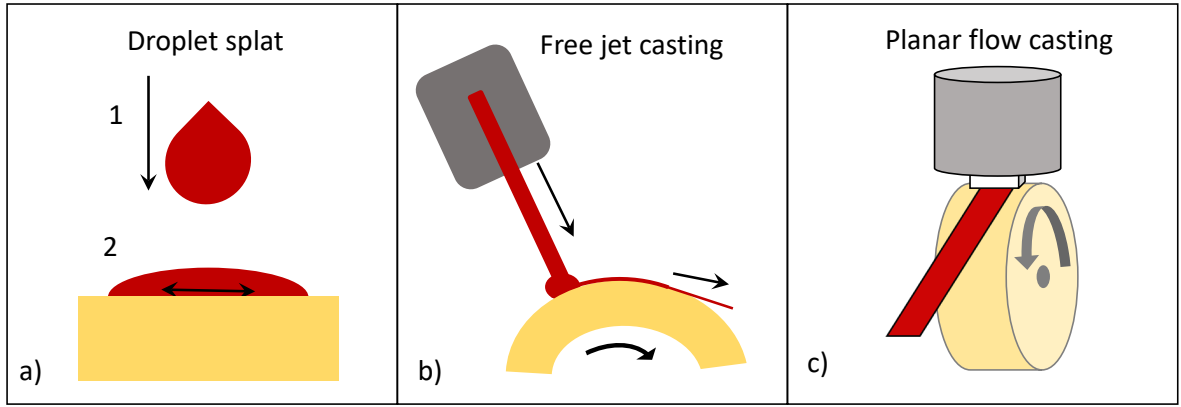

**Figure 1**: Amorphous metal processing techniques. a) Droplet splat method, b) Free jet casting method, c) Planar flow casting (PFC) method.

It is worth noting that a number of fluid instabilities are associated with the free jet process. In particular, capillary vibrations are common since molten metals have high densities, high surface tensions and low viscosities [10]. Nowadays, free jet casting is used mostly commonly to produce intermediate materials that require rapid quenching—but are typically used to create flakes of crystalline rather than amorphous metallic glasses. Some examples are directionally solidified Iron-Boron-Neodymium magnetic flakes and rapidly quenched Aluminum-Iron flakes [11, 12]. The free jet cast foils are then subjected to a sintering process to yield a bonded magnet for the magnetic flakes, and the Aluminum-Iron flakes are forged and subsequently extruded to produce bulk plate or bar materials.

Most of the applications of TMGs, however, require wide and continuous foils. To overcome the narrow widths and flakes obtained via free jet casting, planar flow casting (PFC) was developed. Here, molten metal flows through a wide and narrow slit held in close proximity to the casting wheel; a molten metal liquid puddle bridges the narrow gap spacing between the nozzle and the wheel [13]. The constrained puddle adds some stability to prevent many of the common hydrodynamic instabilities of the free jet process. PFC is most similar to slot coating in terms of the geometry of the nozzle and the puddle configuration. However, in contrast to coating processes where a thin and wide layer of liquid is deposited on a moving substrate and leaves a narrow gap region to be later dried, a solidified ribbon in PFC leaves the liquid puddle region. The physical properties of molten metals are also significantly different from conventional coating fluids and this leads to significantly different process constraints as will be discussed in Section 5.

The PFC method overcomes the width limitation of free jet casting and has become the standard process to produce foils with widths up to ~250 mm. A typical spool of amorphous iron-based foil is shown in Figure 2a with a weight of ~ 800kgs, width of 220mm, thickness of 25 microns, and a coil length of 21,000 m. It is worth noting that TMGs have two distinct surface features that arise from their contact with the wheel or air during solidification. The wheel side of the foil has a matte type finish that tends to have a surface roughness equal to that of the wheel surface. The air side of the foil tends to have a smooth mirror-like finish as shown in Figure 2b.

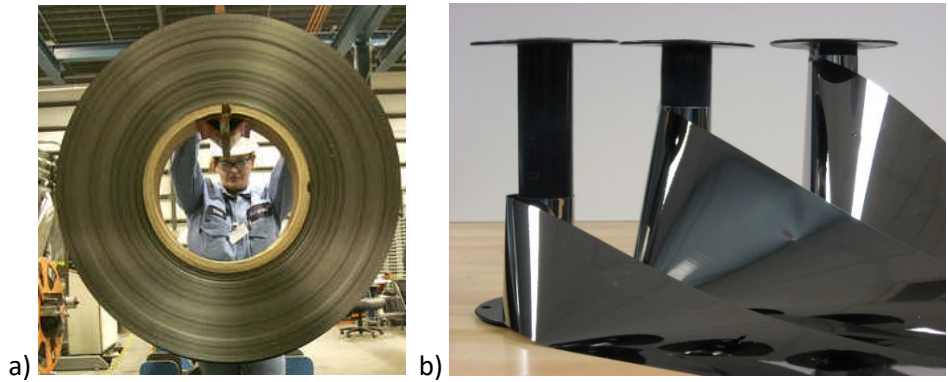

**Figure 2**: a) Typical size of a roll of TMG foil produced by PFC, and b) Mirror-like finish observed on the air side of the foil.

## 3 – Planar Flow Casting (PFC) Process Conditions

The processing conditions used for PFC are common across a wide variety of casting machines and alloys produced. The PFC process, schematically shown in Figure 3, feeds molten metal from a crucible through a nozzle of breadth, $B$, and having a gap, $G$, between the nozzle and casting wheel. Practical constraints on the molten metal dictate that an overpressure, $\Delta P$, be used for fluid delivery. This overpressure, $\Delta P$, consists of the static head of the molten metal and an applied gas pressure that is controlled to be constant as liquid drains from the crucible. The ratio of $B/G > 1$ implies the flow is restricted in the nozzle/wheel gap region, or contact zone, where a puddle of molten metal-- held by surface tension--is formed. With sufficient contact between the wheel and molten puddle, heat is removed from the molten metal and solidification occurs. The solidification front moves at an average velocity, $V$. The wheel moving with linear velocity, $U$, continually removes the solidified product from the contact-zone. The final ribbon thickness, $T$, is primarily controlled by the parameters $B$, $G$, $\Delta P$ and $U$.

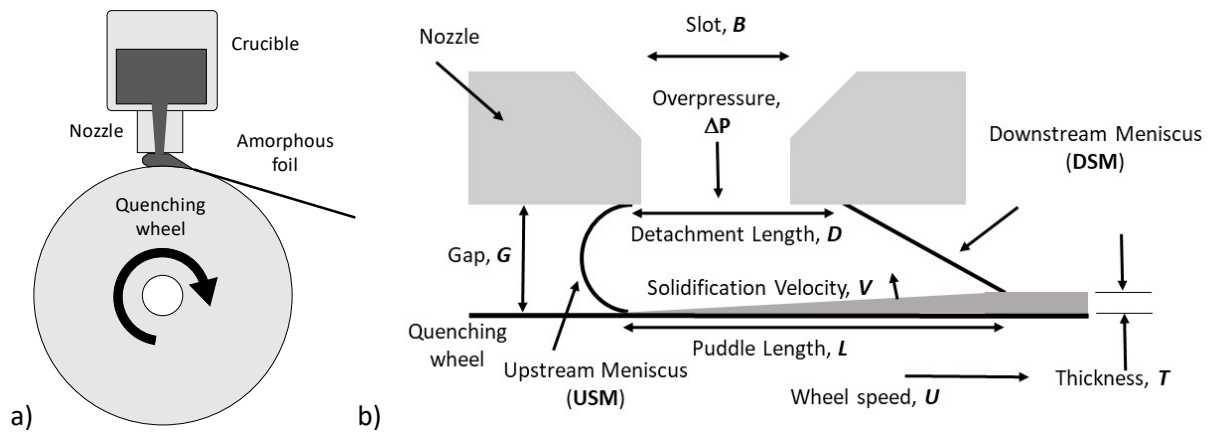

**Figure 3**: a) Schematic of the planar flow casting process and b) details of the contact zone associated within the molten metal puddle region (not to scale).

The process may be run in either batch or continuous modes. In the batch mode, the volume of molten metal in the crucible is finite, and an inert gas, commonly Argon, pressurizes the crucible. To compensate for the decreasing head of liquid, the pressure is increased throughout the cast with the aim being to keep $\Delta P$ constant. In continuous casting mode, molten metal is constantly fed into the crucible which keeps the applied pressure constant. This latter mode requires a two stage process with a holding furnace that supplies molten metal to the crucible.

The casting wheel is most often a copper based alloy because of its high conductivity, and has an internal cooling mechanism. Copper-beryllium was used in some of the initial casting machines because of its combination of high conductivity and hardness. However, the health risks associated with beryllium has largely stopped this practice. The surface of the casting wheel is usually conditioned to remove oxides either before or during the cast to maintain a clean surface. However, when done in continuous mode, this practice can generate a particulate flow of dust particles that can disrupt a cast. The nozzle and crucible materials are typically quartz, graphite boron-nitride, or other ceramics owing to the high melting temperatures required.

Casting is constrained by fundamental process operability limits that enable stable puddle and foil formation. Molten metal must flow into the gap and form a puddle and therefore the applied pressure must overcome the capillary pressure of the upstream meniscus, $\Delta P/(2\sigma/G) > 1$ (visualizations show that the upstream meniscus has a nearly 180 degree contact angle as seen in Figure 4), where $\sigma$ is surface tension. However, if this ratio is much greater than 1 then the upstream meniscus is pushed upstream and out of the puddle region. Inertial forces imparted by the wheel motion to the molten liquid help to balance the forces within the puddle and constrain it. In fact, operability limits are most commonly represented as a function of the Weber number plotted against the applied pressure normalized by the capillary pressure. Figure 4 shows the operability window with experimental high speed video images of the molten metal puddle under stable and failing conditions [14]. The experimental images have a schematic overlaid above to indicate the position of the nozzle (dark grey) and the slot (light grey). From a practical perspective, applied pressures too low will cause the molten metal to freeze within the nozzle. Similarly, sufficiently slow wheel speeds cause the ribbon to freeze within the contact zone. Wheel speeds that are too fast impede solidification and prevent a fully quenched foil from being formed. These basic constraints are useful to identify where the PFC process can operate. However, these limits do not address whether the molten metal can be quenched into an amorphous state, but whether a continuous foil may form.

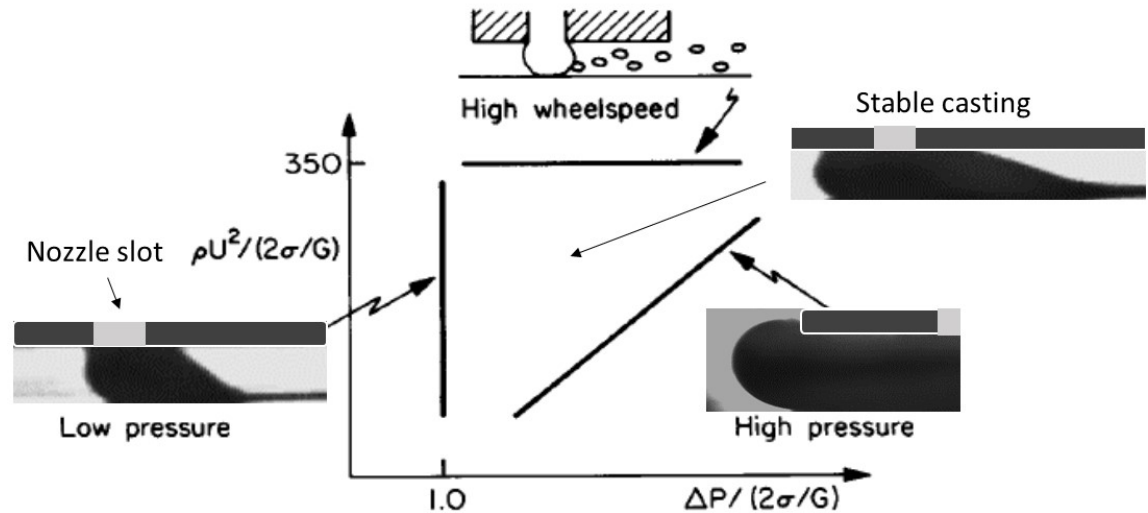

**Figure 4**: Fundamental operability window for the PFC process with experimental PFC puddle images and a schematic overlaid above to indicate the position of the nozzle (dark grey) and the slot (light grey).

It is certainly the case that the PFC process can form rapidly quenched materials that are crystalline. To highlight this, results from two casting machines with different examples of alloy systems will be considered here. The first is the *batch* casting machine at Cornell University that typically produces rapidly quenched Aluminum-Silicon foils. The second is a Metglas commercial *continuous* casting machine used for production of TMGs that are commonly Iron-based amorphous foils. Table 1 shows some of the material properties and casting conditions for these alloys.

**Table 1**: PFC process conditions for the Cornell Aluminum-Silicon (Al-Si)-based crystalline foils, obtained via batch casting, and the Metglas Iron (Fe)-based amorphous foils, obtained via continuous casting.

| Nozzle geometry | Al-Si | Fe-based amorphous | |
|---|---|---|---|
| G | 0.3 - 1.5 mm | 0.15-0.4 mm | Nozzle to wheel gap |
| B | 1.60 - 3.2 mm | 0.4 - 0.6 mm | Nozzle slot breadth |
| W | 25 - 50 mm | 25 - 250 mm | Nozzle/foil width |
| | | | |
| Process variable | | | |
| T | 50 - 350 micons | 15 -75 microns | Range of foil thickness |
| L | 4 - 20 mm | 1 - 5 mm | Typical range of puddle length |
| D | 2 - 10 mm | 0.5 - 2 mm | Typical range of detachment length |
| U | 5 - 15 m/s | 10 - 30 m/s | Wheel speed (linear) |
| Tm | 660 C | 1200 - 1350 C | Molten metal cast temperature |
| | | | |
| Physical properties | | | |
| Density | 2.7 g/cc | 7 - 8.5 g/cc | |
| Viscosity | 1 - 1.3 cP | 4 - 5 cP | |
| Surface tension | 0.86 N/m | 1.2 N/m | |

The Cornell machine data is most useful to examine short-time scale events that happen within a wheel revolution, and the Metglas machine data highlights issues related to continuous production. A recent study focused on these two PFC casting machines and helps to distinguish the physics of amorphous and crystalline processing [15]. It is shown that the ability to cast in PFC is largely insensitive to the morphological nature of the foil being produced and TMGs follow the same basic properties as rapidly quenched crystalline alloys. The alloy composition, the ribbon thickness and critical cooling rates determine the foil microstructure, but surprisingly, the PFC processing conditions do not significantly vary based on the amorphous/crystalline properties of the foil being produced.

The distinction between rapidly quenched TMGs and crystalline foils is not made in the remainder this paper. Rather, fundamental aspects related to PFC processing are highlighted. A simple steady state mass balance around the molten metal puddle region reveals the basic processing relationships since the feed rate ($B*u_{in}$) equals the freeze rate ($L*V$) and also equals the pull rate ($U*T$). Here $u_{in}$ is the inlet flow velocity through the nozzle slot; this is not a specified process parameter since it is the overpressure, $\Delta P$, that is set, but $u_{in}$ can be related back to them through a steady Bernoulli balance as $u_{in} = (G/B)*[2\Delta P/(\rho U)^2]^{1/2}$. Also, the freeze rate involves the puddle length, $L$, as well as the solidification velocity, $V$. More importantly, knowing $U$, $T$, and $B$ allows one to infer the flow rate through the nozzle slot. Similarly, if one can measure $L$, the solidification velocity $V$ can be inferred. The interaction between fluid dynamics and solidification are such that the molten metal puddle length sets the time scale over which solidification occurs or the characteristic residence time within the puddle as $\tau = L/U$. Long puddle lengths imply slow solidification rates and vice versa.

There have been multiple studies that focus on the numerical simulations of the fluid flow and solidification that occurs within the molten metal puddle zone over the past years [16, 17]. Other studies have focused on experimental results where dimensional analysis of the PFC process is used to quantify process conditions [18]. The foil thickness to nozzle gap ratio ($T/G$) is typically 0.1 – 0.2. The nozzle breadth to gap ratio ($B/G$) is typically 1.5 – 5.0 indicating that the gap region is more restrictive than the nozzle in metering the molten metal flow. The general scaling analysis to predict the ribbon thickness is found as,

$$\frac{T}{G} = a \left(\frac{\Delta P}{\rho U^2}\right)^b \quad (1)$$

where $a$ and $b$ are found from experimental casting conditions. Here, $a$ has been reported as 0.5 - 1.0 and $b$ has been reported as 0.25 – 0.5 [19]. The imposed $\Delta P$ may be viewed as synonymous with the flow rate of molten metal. However, the dependence of $T$ on $G$ in Equation (1) reveals the fact that the flow rate is not driven by a pump, since the flow through the gap G in the puddle region affects the flow rate and the ultimate thickness of the foil.

**4 - Defect Formation**

A variety of defects are common in the PFC process and can be found in both crystalline and TMG ribbons. Below find a summary of defects that are frozen into the foil during solidification and are organized by their source. As might be expected, one typical source of defects is from the casting wheel through its vibration and thermal expansion, and the other source is attributed to the dynamics of the liquid puddle itself.

*4.A – Defects Arising from Casting Wheel*

Any type of wheel vibration or eccentricities from out of roundness can get transferred to the foil. The narrow gap, $G$ in Figure 3, meters the flow into the puddle region. Machining variations can be small on the length scale of the wheel diameter but large on the scale of the gap spacing. Plastic deformation of the wheel surface across multiple casting events also contributes to the out-of-roundness. This yields a periodic foil thickness variation on the scale of the wheel circumference due to variation of *G* within a revolution. Figure 5 shows experimental data from the Cornell casting machine that highlights the periodic variation of the foil thickness over 15 cm sections of foil length and a moving average of the thickness measurements over the length scale of the wheel circumference ($C_w$). Here the time between each peak of the ribbon thickness corresponds to that of one wheel revolution.

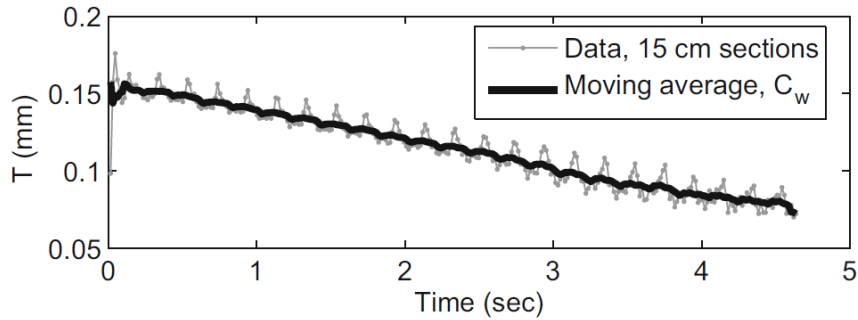

**Figure 5**: Thickness variations observed during the cast length on the Cornell casting machine.

Figure 5 also show a slower and downward trend in thickness that occurs with time during the cast (note the moving average data). This is caused by the wheel expansion as it heats during the cast through contact with the molten metal; the data was obtained from the Cornell batch casting machine that does *not* employ active cooling. This expansion is consistent with the amount of heat transferred into the substrate [20]. The thermal expansion in the *circumferential* direction has the greatest impact on the process and it causes G to slowly decrease during a cast, and this translates into a smaller thickness of the foil over time. Note that even continuous casting machines with active cooling of the quench wheel do experience a transient period where the wheel expands during startup. This thermal expansion during casting is inevitable due to the high heat flow rates through the PFMS process. Variables such as the quench wheel thickness, its internal cooling design, its thermal conductivity, the linear casting speed and many other features determine the amount of thermal expansion that occurs. In a continuous casting mode, this transient period ultimately ends when the heat load and cooling balance to achieve a steady state.

In addition to the variation in gap that affects the lengthwise thickness of the foil, there are also widthwise variations in gap that impart corresponding widthwise variations in foil thickness. Figure 6a shows a cross sectional schematic of the nozzle, the molten metal puddle, and the wheel at the start at the cast. Figure 6b shows the deformation of the casting wheel that arises from its thermal expansion under steady state casting conditions. This expansion is typically symmetric across the width of the foil, with most of the expansion occurring in the center of the rim as shown in in Figure 6b (note that the widthwise expansion of the rim is often a function of the width of foil being produced). The widthwise variation in rim expansion arises since the wheel contact transitions from molten puddle to air in the dashed region shown in Figure 6a. Coupled with the geometrical feature that the rim is wider than the coating, this transition imparts a widthwise component to the heat flow near the edge of the foil, as shown qualitatively in Figure 6d; this yields a widthwise temperature variation and corresponding wheel deformation. In turn, the resulting foil will tend to be thicker on the edges and thinner in the center. The nozzle geometry can be modified to compensate for this effect to maintain a constant gap between the nozzle and the wheel under steady state conditions. Figure 6c shows a contoured nozzle that may be used to accommodate the non-uniform wheel expansion across the width.

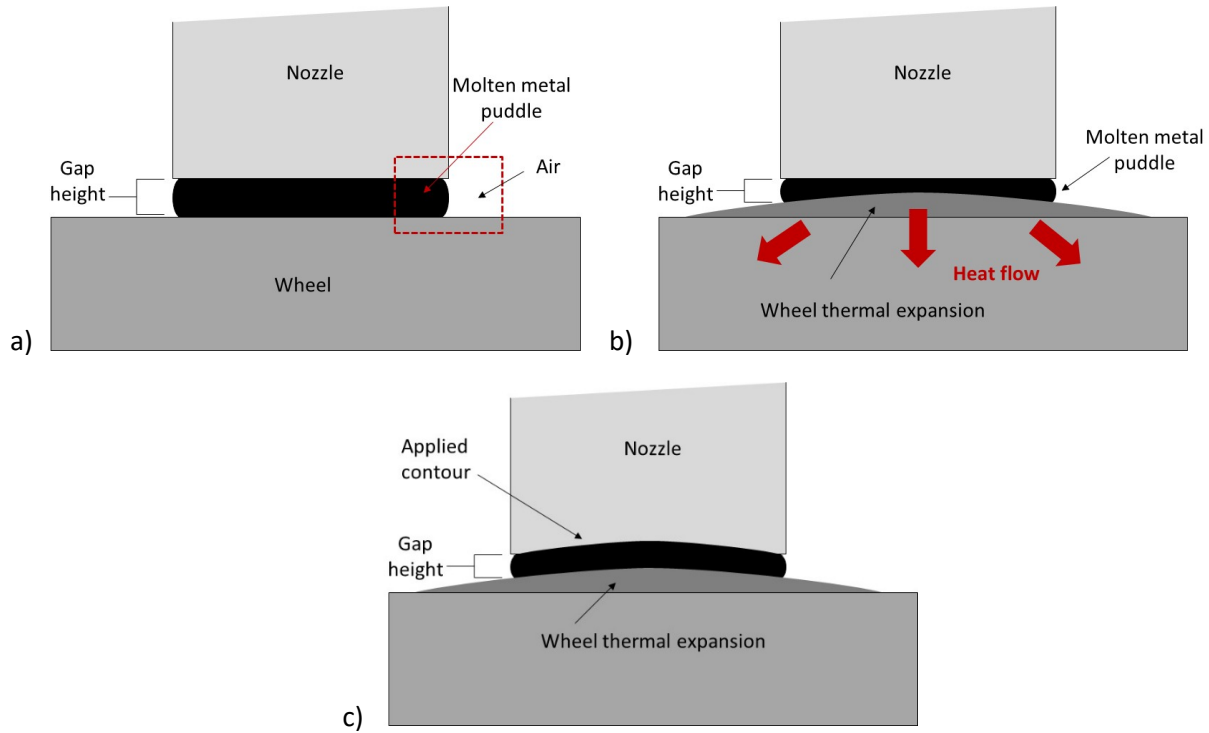

**Figure 6**: Cross sectional view of PFC process. a) There are no thermal deformations of the casting wheel at the start of casting. b) Steady state variations in the gap arise after the casting wheel deforms through dissipation of transient thermal gradients. c) A contoured nozzle shape can match the steady state wheel expansion and produces foil of more uniform thickness across its width. The dashed region in a) shows where the material in contact with the wheel changes from molten metal to air, and this imparts a different thermal wheel expansion near the edge of the puddle compared with its center. Note that the wheel motion is oriented out of the figure.

The casting wheel sees cyclic thermal variations as the wheel comes in contact with the molten metal puddle then cools while leaving the contact zone, with the foil still attached to the rim until it releases downstream. The rim cools further during the revolution only to begin the process again as the puddle region is approached. Figure 7 shows measurements taken during a cast from the Cornell casting wheel via a thermocouple imbedded in the wheel. Here, the time interval between the peaks in the temperature readings correspond to one wheel revolution. This implies that the casting wheel never truly reaches a steady state condition; rather the rim experiences pulses in the heat flux during each revolution.

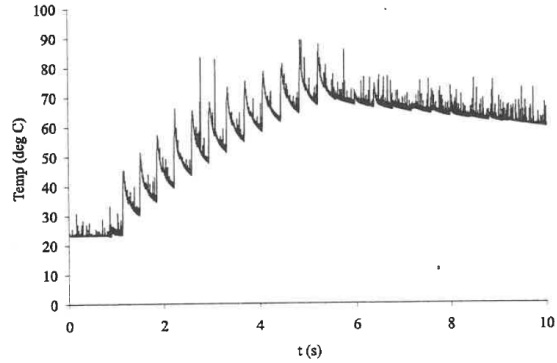

**Figure 7**: Transient temperature measurements taken during a cast from the Cornell casting wheel via a thermocouple imbedded in the wheel.

The data shown in Figure 7 is specific to the casting conditions of the Cornell casting machine where 1 kg of an Aluminum-Silicon alloy was produced in the batch mode. In general, the amplitude of the thermal pulse measured depends on the temperature of the molten metal and the location of the imbedded thermocouples relative to the surface of the casting wheel. However, Figure 7 illustrates an important feature of the PFC process--even under steady casting conditions there are thermal pulses during each wheel revolution that cannot be avoided.

The cast in Figure 7 was limited to 12 wheel revolutions with a casting time of approximately 5 seconds—the temperature was monitored beyond 5 seconds after the cast had completed. When foil is continuously cast, the casting time is on the order of hours and consists of tens of thousands of wheel revolutions. The thermal cycling can lead to fatigue cracks that grow in the wheel substrate. This type of crack propagation leads to mechanical fatigue failures in other metallic production operations. However, few processes see the same type of thermal cycling that the PFC process does, so in this respect the type of thermal cyclic fatigue in PFC is relatively novel. Figure 8 shows how wheel cracks can grow over the length of a cast. As metallic casts are conformal to the casting wheel, the crack depth can be tracked by measuring a corresponding protrusion in the metallic ribbon formed.

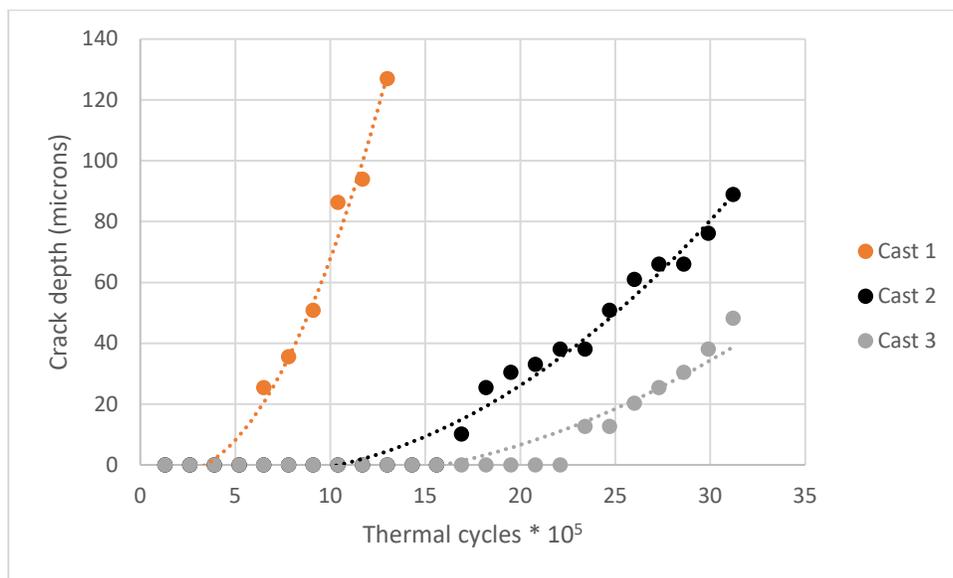

**Figure 8**: Typical thermal cyclic fatigue crack growth over the length of a continuous production cast at Metglas.

The data points shown in Figure 8 corresponds to one measurement per full size spool (c.f. to Figure 2) during an entire production run at Metglas. Many casts exhibit no thermal fatigue cracking and others show a more rapid progression of crack initiation and growth—the sudden onset of cracking is common, and under extreme cases the crack can exponentially grow and fracture the casting wheel. The three casts in Figure 8 represent different generations of casting wheel materials where advances in cast wheel technology minimize the onset of wheel cracking and slow the crack growth rate. Figure 9a shows an optical image of a typical thermal crack observed in a wheel surface and Figure 9b shows a corresponding SEM image of the protuberance feature in the ribbon. The crack propagation occurs in a jagged pattern alternating between axial and circumferential directions on the wheel, as has been reported elsewhere [21].

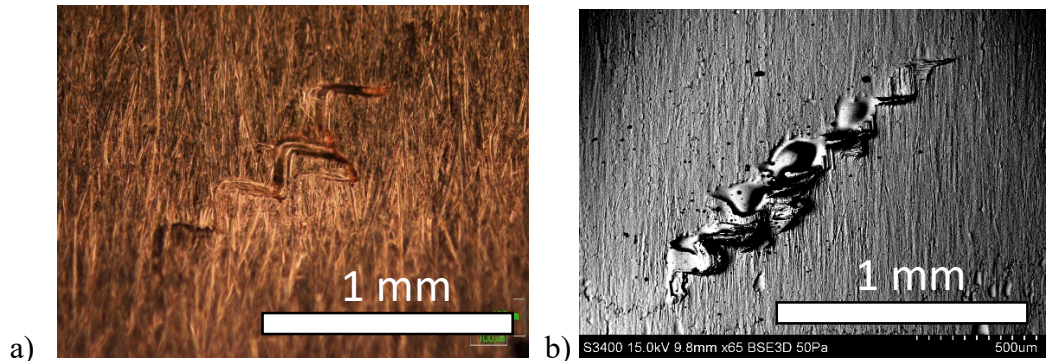

**Figure 9**: a) Crack growth from thermal cyclic fatigue formed in the surface of the casting wheel and b) the crack pattern transferred to the foil during casting.

A crack arising from thermal cyclic fatigue is transferred to the foil during each wheel revolution because the molten metal flows into the wheel crack. In practice, the thermal cyclic cracks on the surface of the casting wheel can be removed by refinishing the surface at the termination of a cast.

*4.B – Flow related defects*

The dynamics of the molten metal puddle itself are also a source of defects. These tend to occur on a time scale associated with fluid dynamics as opposed to time scales imparted by the wheel dynamics. Recall that the high density, high surface tension, and low viscosity of molten metals make the PFC process inherently susceptible to instabilities in ways that conventional viscous coating flows are not.

*Cross-stream defects*

There are a wide range of process conditions that can produce a continuous amorphous foil. However, within these broad operability limits of the process, there are a set of operating parameters that cause the molten metal puddle to freely vibrate at a natural resonant frequency. It has been observed that the upstream meniscus can deform in accordance with classical Rayleigh droplet vibrations [22]. This vibration frequency, *f*, obeys the proportionality

$$f \approx \left(\frac{\sigma}{\rho G^3}\right)^{1/2} \qquad (2)$$

where $\rho$ is the density of the molten metal, *G* is the gap spacing (Figure 3) and $\sigma$ is the molten metal surface tension. Physically this is the ratio of inertial to capillary forces within the puddle and its reciprocal can be viewed as a capillary time scale. Viscous forces are very low in molten metals; thus, there is little vibrational dampening and vibrations can freely resonant. Figure 10 shows a schematic of the upstream meniscus motions observed in the PFC process. These vibrations are correlated to cross stream wave patterns in the cast foil (Figure 11).

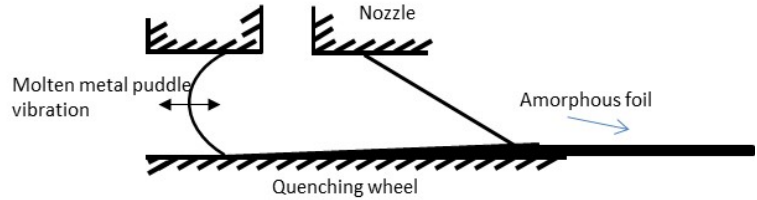

Figure 10: Schematic of the molten metal puddle and upstream meniscus vibration.

A key feature of this vibrational frequency scaling is the nonlinearity of the gap height, and this indicates that control of the gap height is critical. Under specific operating conditions, free puddle vibrations form and are captured in the amorphous foil during processing. Figure 11 shows an image of the free side of the amorphous foil with the puddle vibration captured as a series of scribed lines separated by a wavelength distance $\lambda$. This wavelength can be converted to a frequency by dividing by the linear speed of the quenching wheel as $f = \lambda/U$ where $U$ is the linear wheel speed. The scaling in (2) holds, and a predictive relationship for controlling the wavelength of the pattern in the foil is obtained as

$$\lambda = CU \left(\frac{\rho G^3}{\sigma}\right)^{1/2} \qquad (3)$$

Here $C$ is a constant related to the resonant vibration mode that is experimentally found to be ~0.5.

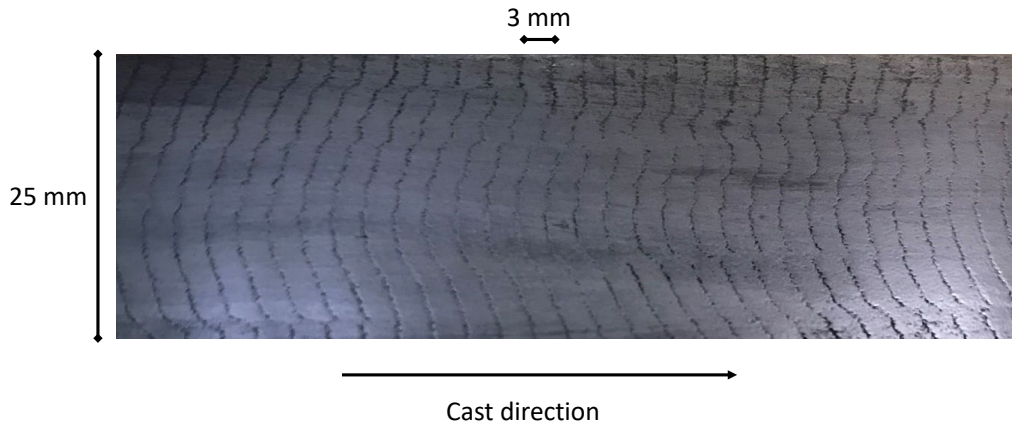

Figure 11: Picture of a typical TMG foil with a cross stream pattern.

This particular cross stream pattern is very robust and can be observed on multiple casting machines across a variety of different alloys in both amorphous and crystalline foils. Figure 12 shows a plot of the measured cross stream frequency against the capillary time scale for an Al-Si crystalline foil (batch cast at Cornell) and an Fe-based amorphous foil (continuously cast at Metglas). The typical wavelength, $\lambda$, is 1-5 mm and when combined with the casting speeds of 10-25 m/s yields a defect frequency of 1-15 kHz. High speed imaging of the molten metal puddle is used to correlate the cross wave pattern in the ribbon to the vibration of the upstream meniscus [22]. A complete mapping of the onset conditions has yet to be completed, but the cross wave pattern can be observed in a subset of conditions in which a ribbon may be formed. Many studies have reported these type of surface features, commonly referred to as herringbone, fish-scale or snake-skin patterns.

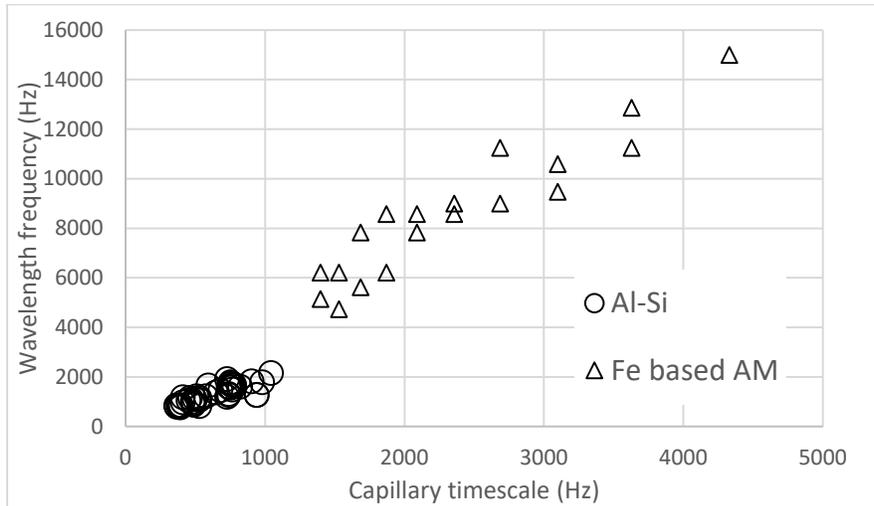

**Figure 12**: Plot of measured cross stream frequency against the capillary time scale for an Al-Si crystalline foil (Cornell) and an Fe-based amorphous foil (Metglas).

*Edge feathers*

The condition of the foil edges is also a critical requirement in most TMG applications. The foil can be cast in widths up to 250mm and then slit into narrower strands. However, slitting costs make this unfavorable and it is more common to cast-to-width, and a uniform foil edge is thus often required. Figure 13 shows a typical desirable foil edge and ones with a feathered appearance. This pattern appears to have the features of a classical Kelvin-Helmholtz instability. Airflow along the edge of a puddle can cause this type of imperfection when the pattern is frozen into the foil edge.

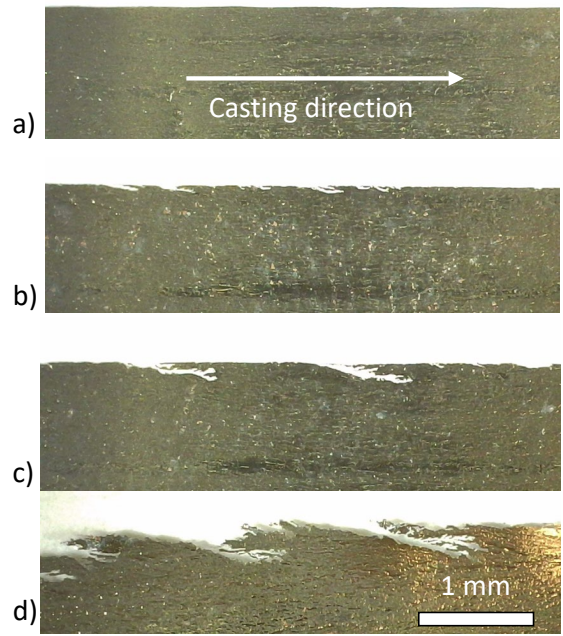

**Figure 13**: Edge conditions of the foil for a) typical desirable cast and b) - d) casts with progressively larger feathered edges. Shown here is a top view of foils near one of its edges.

*Surface roughness / Holes*

The limitation on the maximum casting thickness of a TMG is the critical cooling rate required to maintain an amorphous structure. The limitation on the *minimum* casting thickness is most often the formation of holes. Equation 1 reveals the basic thickness scaling for the PFC process and indicates a reduction in the $G$, reduction in $\Delta P$, or an increase in $U$ will reduce the ribbon thickness. In practice there are tradeoffs when thickness is reduced. A reduction in $G$ needs to be balanced by an increase in $\Delta P$. An increase in $U$ will reduce the ribbon thickness, but also tends to increase the prevalence of holes in the foil. In general, low $G$, high $\Delta P$ and high $U$ promote casting failures and limit TMG productivity.

Holes in the foil are not circular in shape, but rather tend to be extended teardrop shapes oriented in the casting direction (Figure 14). A common cause of holes is copper dust particles that impinge on the puddle (the origin often is due to wheel conditioning during continuous casting; see Section 3), and can occur under normal processing conditions. Additionally, holes can form when the foil thickness approaches the amplitude of the typical surface roughness of the wheel itself.

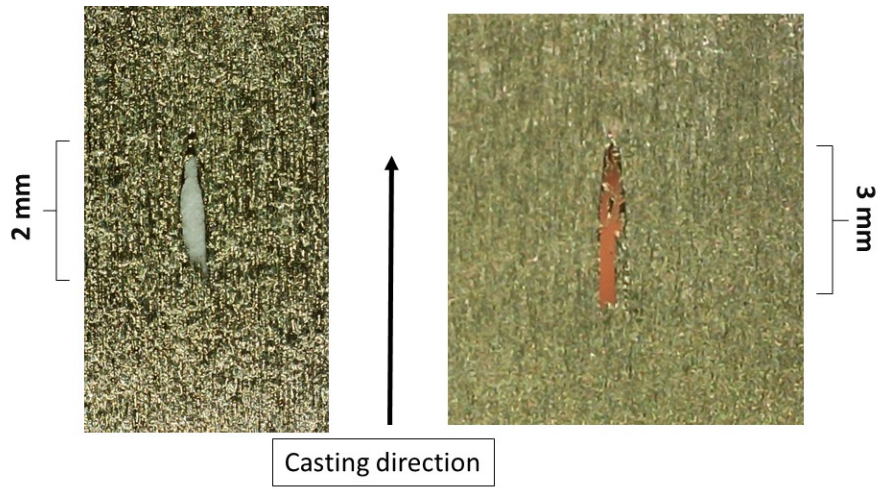

**Figure 14**: Typical surface condition of foil when holes form.

## 5 – Comparison of Planar Flow Casting (PFC) and Conventional Slot Coating

PFC is geometrically most similar to slot coating where liquid is delivered and coated onto a moving substrate. An obvious distinction from PFC is that a substrate to be coated is conveyed into a narrow gap region and is coated with a ribbon of fluid from a coating die [23]. PFC can be viewed as a coating method in which a substrate (the cooling wheel) is coated to produce a solidified thin film that delaminates directly downstream of the coating application point (see Figure 3a). Thus, solidification happens directly in the puddle region itself, while in standard coating processes there is a region downstream of the coating application point in which the coated material is dried to create a solidified product. The casting speeds in PFC are very high compared to conventional coating methods, and this is linked to the constraint that high cooling rates are needed to obtain the metallic glass structures. Thus, in PFC, the maximum casting speed is linked to the speed at which solidification can occur. As in standard coating processes, dynamic wetting of the substrate is required, and while many coating processes are designed to be isothermal, it is not the case in PFC where the start of wetting is concurrent with the initiation of significant heat transfer. In coating processes, the maximum speed at which a coating can occur is typically limited by air entrainment when the dynamic contact angle approaches 180 degrees at the upstream meniscus.

The non-isothermal character of the PFC process is one of the main differences from coating flows, as is the solidification that occurs in the puddle region. In Fe-based TMG production the alloy temperature reduces from ~1300 C down to near ambient temperatures below ~200 C while the foil is in contact with the casting wheel. These extreme temperature gradients make TMGs possible but also cause thermal expansions, cyclic failures of the casting wheel and a handful of other challenges. However, the downstream drying associated with coating flows are not necessary and in some ways this simplifies the process. On the other hand, solidification occurs in the puddle region and the material properties are fixed as the foil exits the puddle. Thus, there are few downstream process variables that can be adjusted to modify the final product with desirable solidified characteristics. In many slot coating processes, the drying rate/conditions can have a significant effect on final product characteristics, and dryers may be quite long. Superficially, the footprint of a PFC machine is much smaller than a coating line in most cases.

Typical coating fluids have significantly different surface tensions compared with molten metals. Molten metals have surface tensions that are 30-40 times larger than for typical aqueous or solvent based coating fluids (coating fluids often have surface tension on the order of 30 dyne/cm). As a result, a capillary meniscus may be maintained with larger gap spacing. In a typical liquid film coating process, the operating gap is typically on the order 200 microns, while that in PFC the gap between the crucible and the wheel (G in Figure 3) can be 10 to 20 times this value. In practice, this larger gap enables simpler nozzle designs where the fine machining associated with slot coating dies are not as critical as those in coating flow methods. This is particularly favorable with PFC where the nozzles are made from ceramic materials that are difficult to machine, and are often replaced. The PFC nozzles need to be heated above molten metal temperatures, and this causes irreversible thermal deformations to the ceramic material upon cooling. Additionally, metal freezes in the nozzle at the cast termination which further prevent the nozzle from being repeatedly used.

In coating processes, the outflow from the bead (equivalent to the puddle in planar flow casting) at the downstream meniscus is essential. At low capillary numbers, $Ca=\mu U/\sigma$, where $\mu$ is viscosity and $U$ is the substrate speed, the Landau–Levich condition sets the radius of curvature, R, of the downstream meniscus for a given coated film thickness. This, in turn, sets the reference pressure in the bead through its curvature, which is imposed on the upstream meniscus. As a result, vacuum is typically needed across the upstream meniscus to enable the meniscus to bridge the gap and meet the imposed dynamic contact angle at the given coating speed. A low *Ca* coating window (operating conditions in which coating is possible) can be made based on geometrical considerations of the bead region and capillary interfaces [24]. At higher flow rates, viscous losses in the bead regions become more relevant, and the Landau-Levich relationship is replaced with equivalent information based on the deformation of the interface due to inertial and viscous effects. In PFC, the downstream interface appears to play a more passive role, and essentially is a flat but angled shape as shown in the images in Figure 4. The pressure dissipation within the puddle is significantly different from any that occurs in slot coating, both due to the inertia-dominated flow and from the effect of the solidification layer on the fluid flow.

The stability window for PFC was discussed in Section 3 where the fundamental process limits are shown. As one traverses the operability region in PFC, the thickness of the foil varies, and the ability of the foil to achieve an amorphous structure is not distinguished. Coating windows discussed above are well known in coating processes, and are largely affected by vacuum applied to the upstream meniscus—and that vacuum is used to help meniscus formation and achieve high coating speeds. In a typical coating window for slot coating, it is straightforward to set the volumetric flow and substrate speed. Thus the thickness of the final wet coating is known a priori, and a coating window applies for a given desired thickness—the coating window is bounded by coating failures, and the condition at which to operate a coating can be determined by minimizing sensitivity to disturbances. Such an approach is largely difficult to do in PFC as the thickness is linked to the location within the operability window. Additionally, vacuum is not typically applied in a PFC process, as the overpressure typically enables a positive pressure in the bead that enables the upstream meniscus to form. Nevertheless, in a preliminary study [25], vacuum has been shown to increase the size of the casting window for PFC in a similar way to that proposed for a low *Ca* coating window [24].

A number of potential foil defects are introduced through the casting wheel itself as it expands and contracts upon contact with molten metal. This gives rise to a periodic variation in foil thickness during each wheel revolution, in part, because the volumetric flow rate is not set via a pump—but rather an

overpressure—the volumetric flow rate is affected if the gap height varies.  Additionally, pressure perturbations can move through a PFC delivery system and impact the volumetric flow rate.  By contrast, in coating processes, positive displacement pumps are used for fluid delivery and this eliminates a feedback mechanism in which the volumetric flow rate is affected by geometrical tolerances in the bead. In the future, it would be interesting to examine the use of a fixed volumetric flow rate to see its effect on foil quality.  One could imaging a plunger being used in the crucible to force the flow in a batch process but this would be difficult to implement in continuous casting.

While there has been some observed coupling between the upstream and downstream interface in Cornell coatings [22, 26], there does not appear to be defects directly attributable to perturbations of the downstream interface.  A number of common defects in PFC were presented in Section 4, and in most of these cases the defect does not necessarily prohibit the use of the TMG in the end application.  The cross-stream defect described previously and commonly found in PFC foils indicates that high frequency inertial-capillary meniscus vibrations can be sustained in a steady casting process.  The foil edge condition may be critical if the foil is used in an as-cast form, but the edge may not be important if the end product is going to be slit from the master coil.  What limits the acceptance of amorphous foil is often application specific. Holes in the foil can be extremely critical or not important at all depending on how the foil will ultimately be used.  Such defects are typically unacceptable in coating flows but only inconveniences in the end application for TMG use.

**6 – Conclusion**

The development of PFC processes has been reviewed and the utility of thin metallic glasses in applications discussed.  We have examined the typical operability range of PFC and the defects that commonly form.  The geometrical similarities between PFC and slot coating process are apparent, and this paper has highlighted both the obvious and more subtle differences in the operability window of such processes.

**7 – Dedication**

This work is dedicated to the memory of Professor Paul H. Steen who passed away in September of 2020.  His innovative fundamental and process research into PFC has transformed casting art into science.  Much of the experimental work presented in this article was performed in Paul's casting laboratory at Cornell University, and the theoretical underpinnings were developed with his students and collaborators.  Paul was inspirational from all perspectives--his technical acumen, his commitment to teaching, and his genuine caring for students and colleagues.  Both of the article co-authors worked closely with Paul--Eric as a PhD student and later in his role at Metglas, and Steve as a research collaborator over the course of many years.  He is sorely missed.

**8 – References**